\documentclass[sigconf,screen]{acmart}

\AtBeginDocument{%
	}

\copyrightyear{2024}
\acmYear{2024}
\setcopyright{rightsretained}
\acmConference[APR '24 ]{2024 ACM/IEEE International Workshop on Automated
	Program Repair}{April 20, 2024}{Lisbon, Portugal}
\acmBooktitle{2024 ACM/IEEE International Workshop on Automated Program
	Repair (APR '24 ), April 20, 2024, Lisbon,
	Portugal}\acmDOI{10.1145/3643788.3648011}
\acmISBN{979-8-4007-0577-9/24/04}

\usepackage{acronym}

\acrodef{AST}{Abstract Syntax Tree}
\acrodef{API}{Ap\-pli\-ca\-tion Programming Interface}
\acrodef{APR}{Automatic Program Repair}
\acrodef{AUG}{API Usage Graph}

\usepackage{xcolor}
\usepackage{graphicx}
\usepackage{xspace}
\usepackage{listings}
\usepackage{tikz}
\usepackage{todonotes}
\usepackage{subcaption}
\usepackage{enumitem}
\usepackage{csquotes}
\usepackage{adjustbox}

\newcommand{\tool}{ASAP-Repair\xspace}

\newcommand{\ovgu}{Otto-von-Guericke University Magdeburg}
\newcommand{\tue}{Eindhoven University of Technology}

\definecolor{ckeywordcolor}{RGB}{127,0,85}
\definecolor{cstringcolor}{RGB}{42,0,255}
\definecolor{ccommentcolor}{RGB}{63,127,95}
\definecolor{addcolor}{RGB}{34,136,51}
\definecolor{delcolor}{RGB}{238,102,119}

\newcommand*\circled[1]{\tikz[baseline=(char.base)]{
		\node[shape=circle,draw,inner sep=0.75pt] (char) {\textbf{#1}};}}

\begin{document}
	\title{\tool{}: API-Specific Automated Program Repair Based on API Usage Graphs}
	
	\settopmatter{authorsperrow=4}
	
	\author{Sebastian Nielebock}
	\email{sebastian.nielebock@ovgu.de}
	\orcid{0000-0002-0147-3526}
	\affiliation{%
		\institution{\ovgu}
		\city{ }
		\country{Germany} 
	}
	
	\author{Paul Blockhaus}
	\email{paul.blockhaus@ovgu.de}
	\orcid{0000-0001-6910-9475}
	\affiliation{%
		\institution{\ovgu}
		\city{ }
		\country{Germany} 
	}
	
	\author{Jacob Krüger}
	\email{j.kruger@tue.nl}
	\orcid{0000-0002-0283-248X}
	\affiliation{%
		\institution{\tue}
		\city{ }
		\country{The Netherlands} 
	}
	
	\author{Frank Ortmeier}
	\email{frank.ortmeier@ovgu.de}
	\orcid{0000-0001-6186-4142}
	\affiliation{%
		\institution{\ovgu}
		\city{ }
		\country{Germany} 
	}
	
	\renewcommand{\shortauthors}{Nielebock et al.}
	
	\begin{abstract}
		Modern software development relies on the reuse of code via \acp{API}. Such reuse relieves developers from learning and developing established algorithms and data structures anew, enabling them to focus on their problem at hand. However, there is also the risk of misusing an \ac{API} due to a lack of understanding or proper documentation. While many techniques target API misuse detection, only limited efforts have been put into automatically repairing \ac{API} misuses. In this paper, we present our advances on our technique \textbf{A}PI-\textbf{S}pecific \textbf{A}utomated \textbf{P}rogram Repair (\tool{}). \tool{} is intended to fix \ac{API} misuses based on \acp{AUG} by leveraging \ac{API} usage templates of state-of-the-art \ac{API} misuse detectors. We demonstrate that \tool{} is in principle applicable on an established \ac{API} misuse dataset. Moreover, we discuss next steps and challenges to evolve \tool{} towards a full-fledged \ac{APR} technique.

	\end{abstract}
	
	\keywords{API Misuses, Automated Program Repair, API Usage Graphs}

	\ccsdesc[500]{Software and its engineering~Error handling and recovery}
	\ccsdesc[300]{Software and its engineering~Software libraries and repositories}
		
	\maketitle

	\section{Introduction}

\looseness=-1
An \acf{API} is the de facto standard for \emph{client developers} to reuse algorithms and code implemented in a library or framework developed by \emph{\ac{API} developers}. An \ac{API} provides a set of \emph{\ac{API} elements} (e.g., methods, fields, data structures). When using these \ac{API} elements, client developers can \emph{misuse} them, for instance, mixing up the order of method calls or calling methods with false parameters~\cite{Amann2019}. If such a misuse causes negative behavior (e.g., software crashes, performance issues), we refer to it as \emph{\ac{API} misuse}.

\looseness=-1
In the past, many techniques for detecting \ac{API} misuses have been developed~\cite{Ammons2002,Weimer2005,Zhong2009,Amann2019,Nielebock2020,Kang2021}. Commonly, these techniques infer or mine likely correct \emph{template usages} of the \ac{API} (e.g., patterns or change rules) whose violations are reported as misuses. While limited, such techniques provide a profound way to specifically detect \ac{API} misuses. Importantly, they \emph{do not require dedicated tests to detect and localize misuses}, and they provide a patch via the template usage with which the misuse has been detected. A natural way to advance research is to leverage these templates for automated repair. 

In this paper, we introduce the idea of an \acf{APR} technique for \ac{API} misuses to which we refer to as \textbf{A}PI-\textbf{S}pecific \textbf{A}utomated \textbf{P}rogram Repair (\tool{}). \tool{} builds on a graph-based structure representing \ac{API} usages named \acfp{AUG}, which has been introduced by \citet{Amann2019a}. Particularly, we leverage different template usages represented as \acp{AUG} (i.e., patterns and change rules) to repair \ac{API} misuses. While currently limited to repairing misuses in \acp{AUG} and not in code, \tool{}'s major benefit compared to state-of-the-art \ac{APR} techniques is that it \emph{does not require test cases or the execution of code to localize misuses}. We demonstrate its in-principle applicability by applying \tool{} on the real-world \ac{API} misuse dataset MUBench~\cite{Amann2016}. Moreover, we discuss necessary steps and challenges to evaluate and to compare \tool{} with state-of-the-art \ac{APR} techniques. We publish \tool{} and all artifacts related to this paper in a publicly available repository.\footnote{\url{https://doi.org/10.5281/zenodo.10527304}}

	\section{API Usage and API Misuse Detection}

\begin{figure}
	\begin{subfigure}[b]{0.49\linewidth}
		\includegraphics[width=\linewidth, trim={38 38 38 38},clip]{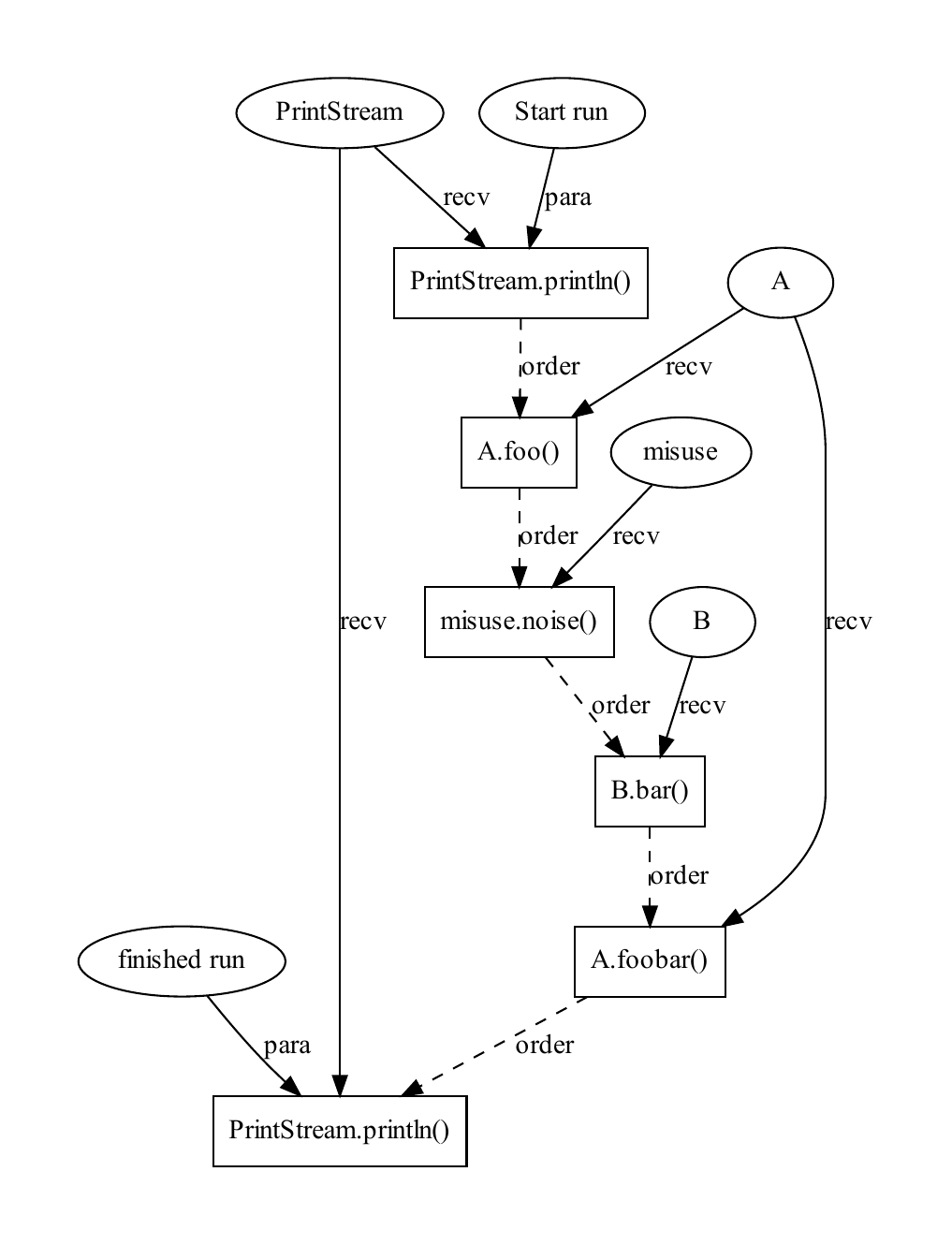}
		\caption{\ac{AUG} of the sample code.\label{fig:aug}}
	\end{subfigure}
	\begin{subfigure}[b]{0.49\linewidth}
		\includegraphics[width=\linewidth,trim={48 48 48 48},clip]{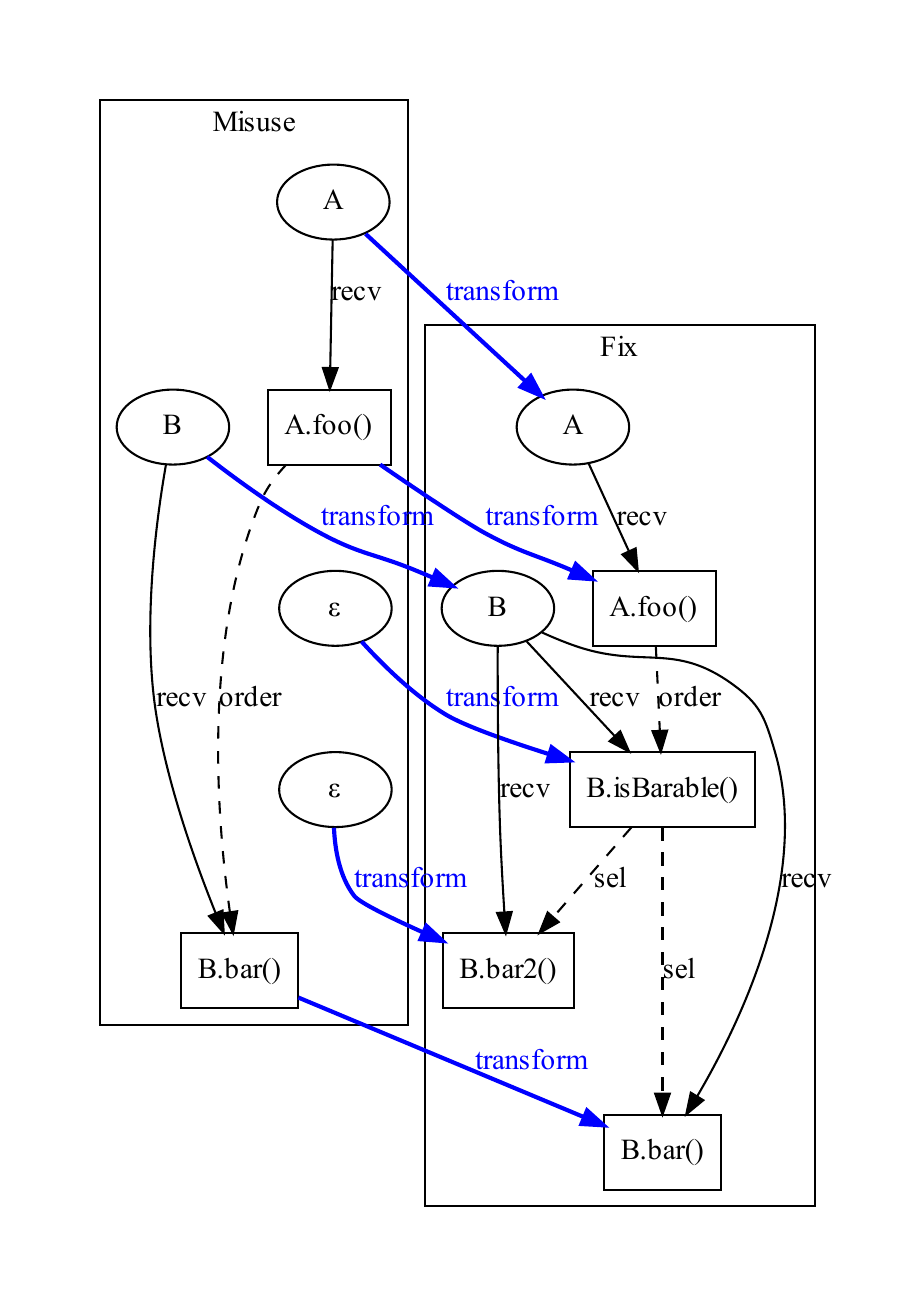}
		\caption{Change rule of the misuse fix.\label{fig:change_rule}}
	\end{subfigure}
	\vspace*{-2ex}
	\caption{\ac{AUG} (left) and change rule (right) for the example code in \autoref{fig:code}.}
	\label{fig:example}
	\vspace*{-2ex}
\end{figure}

\paragraph*{API Usage Graphs} \citet{Amann2019a} have developed a graph-based structure to describe \ac{API} usages in Java with the goal to improve the precision of \ac{API} misuse detection. For representational purposes, we introduce \acp{AUG} with a fictive fix\footnote{Please refer to our replication package for real examples.} of an \ac{API} misuse depicted in \autoref{fig:code} and its \ac{AUG} representation before this fix (cf. \autoref{fig:aug}). This fix adds a validation for the method call \lstinline[basicstyle=\normalsize\ttfamily]|B.bar()| in line 9 by checking the condition \lstinline[basicstyle=\normalsize\ttfamily]|B.isBarable()|. In case this resolves to \lstinline[basicstyle=\normalsize\ttfamily]|false|, an alternative call to \lstinline[basicstyle=\normalsize\ttfamily]|B.bar2()| is required.


\acp{AUG} are directed, labeled, acyclic multigraphs consisting of different node (i.e., action and data) and edge types (i.e., data and control flow). \emph{Action nodes} are depicted as rectangles representing method calls (e.g., \lstinline[basicstyle=\normalsize\ttfamily]|A.foo()|) or control structures, while ellipses visualize \emph{data nodes}, such as constants and objects (e.g., \lstinline[basicstyle=\normalsize\ttfamily]|A|). \emph{Data flow} is represented with solid edges connecting parameters to methods (i.e., \texttt{para}) or objects to their object methods (i.e., \texttt{recv}). \emph{Control flow} is displayed via dashed edges that indicate the order of action nodes (i.e., \texttt{order}) or the selection after \lstinline[basicstyle=\normalsize\ttfamily]|if|-conditions (i.e., \texttt{sel}).
\acp{AUG} only represent the \ac{API} usage of a single method declaration (e.g., \lstinline[basicstyle=\normalsize\ttfamily]|method()|), and thus are restricted to the intra-procedural level. However, they have been demonstrated to improve \ac{API} misuse detection and were enhanced as well as used by different researchers~\cite{Amann2019a,Nielebock2020,Kang2021}; making them a valid basis for an \ac{APR} technique.

\begin{lstlisting}[float,belowskip=-3ex,caption={Sample code for an API misuse and fix.},label=fig:code]
import A;
import B;
public class misuse {
    public void method() {
        System.out.println("Start run");
        A.foo();
        noise();
-        B.bar();
+        if(B.isBarable()){
+            B.bar();
+        } else{
+            B.bar2();
+        }
        A.foobar();
        System.out.println("finished run");
    }
}
\end{lstlisting}

\paragraph*{Frequent API Usage Patterns} Researchers have proposed a set of \ac{API} misuse detectors, many of them applying frequent pattern mining~\cite{Ammons2002,Weimer2005,Zhong2009,Amann2019,Kang2021}. Their conjecture is that the way \acp{API} are frequently used also represents their correct usages. Thus, they conduct pattern mining to infer \ac{API} usage patterns and compare those with actual usages of that \ac{API}. If a usage violates this pattern, it is denoted an \ac{API} misuse. However, recent work showed that such misuse detectors typically suffer from a high false positive rate (i.e., a huge number of correct usages reported as misuses)~\cite{LeGoues2009,Amann2019}, and thus research has focused on improving the precision of misuse detectors. In the following, we assume that a pattern in the form of an \ac{AUG} has been found that can correctly detect an \ac{API} misuse.

\paragraph*{API Change Rules} Another idea leverages information from already fixed misuses by extracting \ac{API} code changes using version control~\cite{Nielebock2020,Nielebock2021b}. In detail, such techniques infer so-called correction or change rules, which represent the essential changes needed to fix an \ac{API} misuse. A rule has the form $m \rightarrow f$ where $m$ and $f$ are the misuse and fixed (sub-)\acp{AUG} of the change, respectively. We show the rule for the fix in \autoref{fig:code} in \autoref{fig:change_rule}. To match both graphs, heuristics are applied, which we also apply for \tool{} (cf. \autoref{sec:process}). A rule consists of the subgraphs describing the misuse and its fix as well as \texttt{\color{blue}transform}-edges representing how nodes are transformed into their respective fix. Additions are symbolized by special $\epsilon$-nodes indicating \enquote{holes} in the misuse graph. Similarly, deletions are represented by using $\epsilon$-nodes in the fixed \ac{AUG}.

To detect \ac{API} misuses, existing techniques measure the similarity\footnote{Originally, this was called \enquote{distance,} which is mathematically restricted. Thus, we use the more general term \enquote{similarity.}} $sim$ of a candidate \ac{API} usage $u$ and both subgraphs $m$ and $f$ of the rule. A usage $u$ is reported as a misuse if  $sim(m,u)>sim(f,u)$ holds, meaning that the usage is more similar to the misuse part than to the fixed part. Results indicate that such a technique can achieve high precision, but suffers from a very low recall~\cite{Nielebock2022}.

	\section{Process of \tool{}\label{sec:process}}

\begin{figure*}
	\includegraphics[width=\textwidth, trim={0 230 210 0},clip]{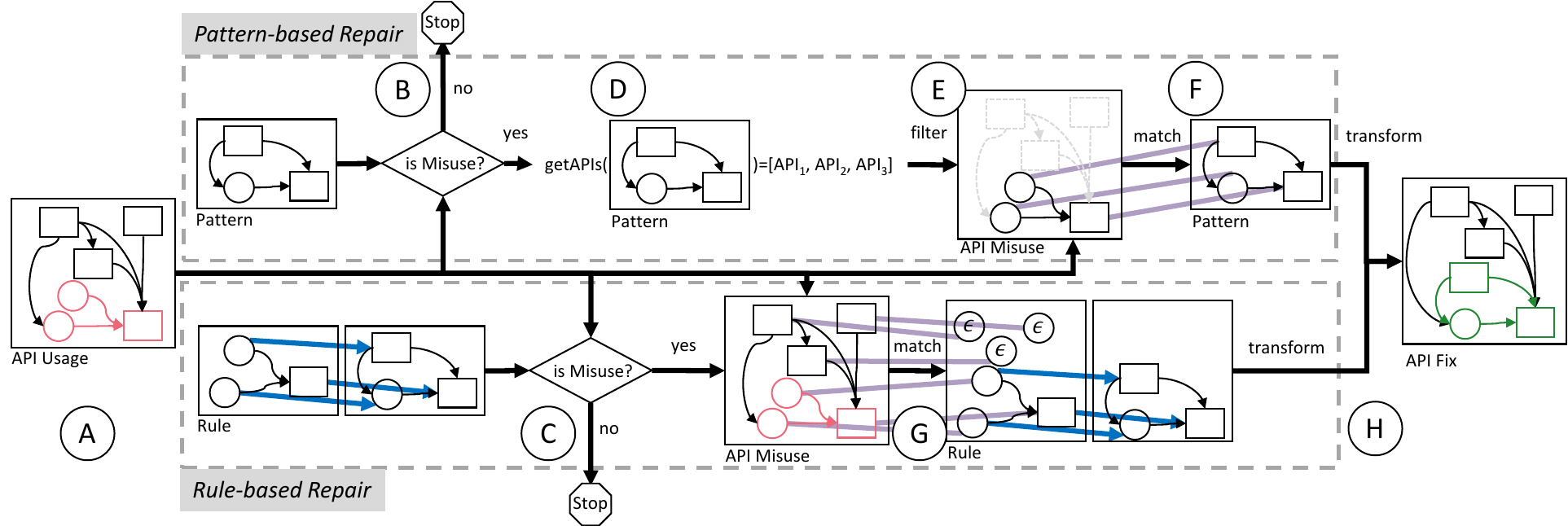}
	\vspace*{-4ex}
	\caption{Concept of Pattern- and Rule-based API-specific Automated Program Repair\label{fig:repair_concept}}
	\vspace*{-2ex}
\end{figure*}

We depict \tool{}`s concept in \autoref{fig:repair_concept}. It fixes \ac{API} misuses in the form of \acp{AUG} using \emph{patterns} (i.e., as \acp{AUG}) or \emph{change rules}. Both variants start with an \ac{API} usage ({\footnotesize\circled{A}}), which is transformed into its respective \ac{AUG} followed by the misuse detection. For the pattern-based version ({\footnotesize\circled{B}}), we can apply the violation-based technique by \citet{Amann2019a}. When using change rules ({\footnotesize\circled{C}}), we can use the similarity-based technique~\cite{Nielebock2021b,Nielebock2022}. If a misuse is detected, \tool{} has different steps to match the nodes of the misuse \ac{AUG} with those of the template \ac{AUG} (i.e., pattern or change rule) through which the misuse was detected. Using this matching, we identify which nodes have to be changed (i.e., add, delete, update).

\paragraph*{Matching Heuristic} Matching graphs refers to the subgraph isomorphism problem known to be NP-complete~\cite{Grohe2020}. Thus, we adapted the strategy to produce change rules~\cite{Nielebock2020,Nielebock2022} for graph matching. In detail, we applied the Kuhn-Munkres algorithm~\cite{Munkres1957} to find a heuristic solution for the matching. This means that we create a bipartite graph based on the two \acp{AUG} (e.g., $aug_A$ and $aug_B$), with one partition containing the nodes of $aug_A$ and the other the nodes of $aug_B$. Then, we equalize the cardinalities of both partitions by adding special $\epsilon$-nodes. These describe the addition and deletion of nodes (e.g., a node changed to an $\epsilon$-node represents a deletion) within the matching. Finally, we add edges between the nodes of both partitions, which we label with the costs to transform the respective node into the other (i.e., the number of node relabelings as well as adding, deleting, and relabeling incoming and outgoing edges). The Kuhn-Munkres algorithm finds a matching that minimizes the overall costs. Note that this matching can be invalid, since it ignores the order of nodes and multiple matchings are possible.

\paragraph*{Pattern-Based Matching} In case we repair an \ac{API} misuse based on patterns, we cannot directly match the misuse with the pattern \ac{AUG}. The reason is that the pattern is typically smaller (in terms of number of nodes) than the misuse. Thus, a matching would indicate that every node not part of the pattern must be deleted. We avoid this issue by conducting a preprocessing of the misuse to find and match only those nodes that relate to the pattern. Particularly, we leverage that \acp{AUG} contain information on the \ac{API} type of single nodes (e.g., \lstinline[basicstyle=\normalsize\ttfamily]|java.util.List|). Then, we consider only those nodes of the misuse for matching that have the same \ac{API} type as the nodes in the pattern. Thus, we determine the \ac{API} types present in the pattern \ac{AUG} ({\footnotesize\circled{D}}) and filter nodes from the misuse \ac{AUG} to extract a subgraph, consisting only of nodes of these types ({\footnotesize\circled{E}}). Then, we match only this subgraph to the pattern ({\footnotesize\circled{F}}).

\paragraph*{Rule-Based Matching} Matching an \ac{API} misuse with a change rule is simpler, since such rules describe exactly which nodes have to be changed. In detail, we match the misuse \ac{AUG} with the misuse part \ac{AUG} of a change rule ({\footnotesize\circled{G}}). If the rule contains an addition of nodes (i.e., the misuse part contains $\epsilon$-nodes), we temporarily disregard these from the matching procedure. Then the matching can have three possible cases:
\begin{enumerate}[nosep,leftmargin=*]
	\item A node from the misuse \ac{AUG} is matched to a non-$\epsilon$-node of the misuse part \ac{AUG}. This indicates that this node has to be either deleted or updated  (i.e., depending on whether this node is connected via a \texttt{\color{blue}transform}-edge to an $\epsilon$-node or not).
	\item A node from the misuse \ac{AUG} is matched to a generated $\epsilon$-node in the misuse part (i.e., due to the cardinality equalization step of the matching). Then, this node will not be part of the transformation, since it cannot be matched to the change represented by the rule.
	\item For every disregarded $\epsilon$-node in the misuse part \ac{AUG} of the rule, we add a respective $\epsilon$-node in the misuse \ac{AUG} as well. We match those to the counterparts in the misuse part \ac{AUG} to indicate that nodes have to be added according to the fixed part \ac{AUG} of the change rule. 
\end{enumerate}
This way, we obtain a triple matching between misuse \ac{AUG}, misuse part \ac{AUG}, and fix part \ac{AUG}. Note that nodes of the misuse \ac{AUG} falling under case (2) are not part of the final repair step.

\paragraph*{AUG-Based Repair} In the final step of \tool{}, we transform the misuse graph into the fixed version ({\footnotesize\circled{H}}). We refer to the changes indicated by the respective matching, either pattern-based or rule-based, as \emph{corrections}. We distinguish between three cases of transformations/corrections:
\begin{enumerate}[nosep,leftmargin=*]
	\item \emph{A non-$\epsilon$-node from the misuse is matched to a non-$\epsilon$-node in the correction}: Then, \tool{} \emph{updates} the node by updating its label, node type, as well as adding, deleting, updating the respective incoming and outgoing edges.
	\item \emph{An $\epsilon$-node from the misuse is matched to a non-$\epsilon$-node in the correction}: \tool{} \emph{adds} the non-$\epsilon$-node to the \ac{AUG}, which also contains the addition of the respective edges represented by the correction.
	\item \emph{A non-$\epsilon$-node from the misuse is matched to an $\epsilon$-node in the correction}: \tool{} \emph{removes} the non-$\epsilon$-node from the \ac{AUG}, which also contains the deletion of the respective incoming and outgoing edges of this node.
\end{enumerate}
Note that \tool{} keeps nodes unmatched (i.e., filtered out in the pattern-based repair or case (2) in the rule-based repair), except by changing incoming or outgoing edges from previously changed neighbor nodes. Since the Kuhn-Munkres algorithm does not always produce a valid matching, it may happen that the transformed \ac{AUG} contains cycles, which results in an invalid \ac{AUG}.  We tackle this problem by conducting a graph cycle check.
If the transformed \ac{AUG} contains cycles, we retry the transformation by testing another matching, which we obtain based on an internal data structure of the Kuhn-Munkres algorithm. In case none of the possible matchings produces a valid \ac{AUG}, no repaired \ac{AUG} is generated.

	\section{Preliminary Results}

\looseness=-1
We conducted a preliminary evaluation of \tool{} by applying it to MUBench,\footnote{\label{foot:mubench}\url{https://github.com/stg-tud/MUBench/}} a dataset consisting of real \ac{API} misuses that has been constructed by \citet{Amann2016}. More precisely, we used the subset of 116 misuses provided in the replication package by \citet{Nielebock2021b}.
In our evaluation, we conducted a sanity check on whether \tool{} is applicable for real misuses. That means, for every single \ac{API} misuse, we used its fix as a pattern and the changes of the fixing commit as the basis for the change rule. This way, we validated \tool{} in an ideal situation, in which a perfectly matched pattern and change rule are found.  Therefore, we cannot draw general conclusions about the applicability of \tool{} in practice or in comparison to state-of-the-art \ac{APR} techniques, which are subject to our future work.
 
\begin{table}
	\caption{Results of repairing 116 misuses form MUBench:\\ D - \#generated \textbf{d}ata structures for repair (i.e., pattern \ac{AUG} or change rule), C - \#\textbf{c}reated, V - \#\textbf{v}alid, and U - \#\textbf{u}nique fixes.}
	\label{tab:results}
	\vspace*{-2ex}
	\begin{adjustbox}{width=\linewidth}
		\begin{tabular}{lrrrr}
			\toprule & \textbf{D (\%)} & \textbf{C (\%)} & \textbf{V (\%)} & \textbf{U (\%)} \\
			\midrule
			\textbf{pattern-based} & 110 (94.8\%) & 61 (52.6\%) & 34 (29.3\%) & 11 (9.5\%)\\ 
			\textbf{rule-based} & 86 (74.1\%) & 38 (32.8\%) & 27 (23.3\%) & 4 (3.4\%)\\
			\bottomrule
		\end{tabular}
	\end{adjustbox}
	\vspace*{-2ex}
\end{table}

For each example in the dataset, we downloaded the code version directly before and after a fixing commit, representing the misuse and fixed version, respectively, and used them to mimic the pattern and change rule. For each repair, we set a timeout of five minutes. Then, the first author checked how many repaired \acp{AUG} \tool{} generated and manually validated each fix using an \ac{AUG} comparison technique to decide whether the \ac{AUG} of the fixed version and the generated fix were semantically equal.
In \autoref{tab:results}, we summarize the results. We observed that \tool{} produced more valid pattern-based fixes than rule-based ones (i.e., $34$ vs $27$) with more unique fixes (i.e., $11$ vs. $4$). However, the rule-based repair obtained a larger proportion of valid fixes within the generated fixes (i.e., $27/38\approx71\%$ vs. $34/61\approx55.7\%$), \emph{indicating a better precision of the rule-based repair}. We also analyzed the reasons why the other repairs could not be generated or were invalid. For the pattern-based repair, major issues were that fixed \acp{AUG} contained cycles (36 cases), invalid edges in the fixed \ac{AUG} (16 cases), or timeouts during the repair (11 cases). For the rule-based repair, we encountered timeouts and out-of-memory exceptions (59 cases) as well as invalid edges in the fixed \acp{AUG} (11 cases). Moreover, we found 16 cases in which MUBench had a false misuse description causing false or no repair at all. Three of these false descriptions caused an unsuccessful pattern-based \ac{AUG} generation. For more details, we refer to our replication package.

While still limited, these results are promising to further improve and apply \tool{} as a full-fledged \ac{APR}-technique. It provides a framework to synthesize results from pattern-based and rule-based misuse detectors in a single \ac{APR} technique. The main challenge lies in improving the matching efficiency as well as prohibiting the construction of invalid fixed \acp{AUG}.

	\section{Conclusion and Prospects}

We demonstrated that \ac{AUG}-based repair is possible for real \ac{API} misuses. To obtain a valid \ac{APR} technique, we will continue addressing the following prospects.

\paragraph*{From \acp{AUG} to Code} While \acp{AUG} represent a good visual way to describe required \ac{API} fixes, a more practical approach is a full-fledged technique producing repaired source code. We can imagine two possible ways to achieve this: (1) We directly transform an \ac{AUG} into its respective source code. This requires a valid specification on how to transform code into an \ac{AUG}, which is only implicitly given by the implementation\textsuperscript{\ref{foot:mubench}} of MUBench~\cite{Amann2019a}. Moreover, we need the back-transformation from \acp{AUG} to code. (2) We use the matching to perform the respective code transformations. In detail, we need to define for each possible node and edge transformation in the \ac{AUG} a proper abstract-syntax-tree transformation. 

\paragraph*{Representative Misuse Datasets} While MUBench~\cite{Amann2016} is a valid misuse dataset, its representativeness is debatable, as many entries represent essentially the same misuse. Thus, other misuse datasets, such as AU500~\cite{Kang2021} with its manually labeled misuses, must be used, too. However, these are limited in their size, which, in turn, limits external validity. Thus, larger validation datasets must be build.

\paragraph*{Comparison to State-of-the-Art \ac{APR} Techniques} Many \ac{APR} techniques have been developed in the past~\cite{Monperrus2018}, to which we have to compare \tool{}. While we validated \tool{} manually and statically, we also have to validate whether the code fixes imply a correct dynamic behavior. This requires the execution of the code using test cases. A good starting point is the APIARTy framework by \citet{Kechagia2021}, which analyzes state-of-the-art \ac{APR} techniques on \ac{API} misuses that were validated via tests. However, we noticed that some misuses are not replicable, since repositories became outdated. Thus, their build process does not work anymore. So, the underlying misuse data and build commands have to be updated manually to ensure a valid evaluation and comparison.

	\balance
	\bibliographystyle{ACM-Reference-Format}
	\bibliography{publishers,fullBib,MYabrv,asap}

\end{document}